\documentclass[pra,aps,amsfonts,amsmath,superscriptaddress,%
  twocolumn,showpacs,floatfix]{revtex4}
\usepackage{amsfonts}
\usepackage{amsmath}
\usepackage{bm}
\usepackage{epsfig}

\renewcommand{\Im}{\mathrm{Im}\,}
\newcommand{\Ref}[1]{Ref.\@ \cite{#1}}

\newcommand{\Eq}[1]{Eq.\@ (\ref{#1})}
\newcommand{\Eqs}[1]{Eqs.\@ (\ref{#1})}
\newcommand{\Tab}[1]{Table \ref{#1}}
\newcommand{\Fig}[1]{Fig.\@ \ref{#1}}
\newcommand{\Figs}[1]{Figs.\@ \ref{#1}}
\newcommand{\Sec}[1]{Sect.\@ \ref{#1}}
\newcommand{\vek}[1]{\bm{\mathrm{#1}}}
\newcommand{\rp}{r^\prime}
\newcommand{\sigmap}{\sigma^\prime}
\newcommand{\rpp}{r^{\prime\prime}}
\newcommand{\vv}{\vek{v}}
\newcommand{\pv}{\vek{p}}
\newcommand{\rv}{\vek{r}}
\newcommand{\rvp}{\vek{r}^\prime}

\newcommand{\pvp}{\vek{p}^{\prime}}
\newcommand{\nablav}{\vek{\nabla}}
\newcommand{\xiv}{\vek{\xi}}
\newcommand{\llangle}{\langle\langle}
\newcommand{\rrangle}{\rangle\rangle}
\newcommand{\muloc}{\mu_{\mathit{loc}}}
\newcommand{\Ekin}{E_{\mathit{kin}}}
\newcommand{\Epot}{E_{\mathit{pot}}}
\newcommand{\Eint}{E_{\mathit{int}}}
\newcommand{\Vext}{V_{\mathit{ext}}}
\begin{document}
\title{Temperature and finite-size effects in collective modes of
superfluid Fermi gases}
\author{M. Grasso}
\affiliation{Dipartimento di Fisica ed Astronomia and INFN, Via Santa
  Sofia 64, I-95123 Catania, Italy}
\affiliation{Institut de Physique Nucl\'eaire, Universit\'e Paris-Sud,
  IN2P3-CNRS, 91406 Orsay Cedex, France}
\author{E. Khan}
\affiliation{Institut de Physique Nucl\'eaire, Universit\'e Paris-Sud,
  IN2P3-CNRS, 91406 Orsay Cedex, France}
\author{M. Urban}
\affiliation{Institut de Physique Nucl\'eaire, Universit\'e Paris-Sud,
  IN2P3-CNRS,
91406 Orsay Cedex, France}
\begin{abstract}
  We study the effects of superfluidity on the monopole and quadrupole
  collective excitations of a dilute ultra-cold Fermi gas with an
  attractive interatomic interaction. The system is treated fully
  microscopically within the Bogoliubov-de Gennes and quasiparticle
  random-phase approximation methods. The dependence on the
  temperature and on the trap frequency is analyzed and systematic
  comparisons with the corresponding hydrodynamic predictions are
  presented in order to study the limits of validity of the
  semiclassical approach.
\end{abstract}
\pacs{03.75.Ss,21.60.Jz}
\maketitle

\section{Introduction}
Dilute gases of alkaline fermionic and bosonic atoms are superfluid at
very low temperature: Bose-Einstein condensates have been obtained in
the case of bosonic atoms \cite{bec}, while condensation of molecules
(made out of two atoms) has been observed in the case of fermionic
atoms \cite{fer}. For fermionic atoms in the weakly interacting regime
($k_F |a|\ll 1$, where $k_F$ is the Fermi momentum and $a$ the s-wave
scattering length) BCS superfluidity is expected in the case of
attractive interatomic interaction ($a<0$).

A striking experimental evidence for BCS superfluidity is still
missing, even though various signals which would be coherent with a
superfluid behavior have been observed in some experiments: the
anisotropic expansion of the gas after releasing it from the trap
\cite{expa}, the measurement of the gap \cite{gap}, the measurement of
the frequencies and damping rates of the breathing modes \cite{modes}.

However, the gap has been actually measured only in the strongly
interacting regime and no experimental values exist for the weakly
interacting case. The anisotropic expansion on the one hand and the
frequencies of the breathing modes on the other hand can be predicted
within a hydrodynamic approach for a superfluid gas
\cite{Menotti,Stringari,baranov}. In both cases the experimental
observations agree very well with the hydrodynamic predictions, and
this could actually be considered as an evidence for
superfluidity. However, the predictions for a superfluid gas are the
same as those for a normal gas in the presence of collisions. It is
true that at the very low temperatures achieved in these experiments
Pauli principle is expected to inhibit collisions.  However, the
experimental measurements have been performed during the expansion of
the gas after releasing it from the trap.  In such a situation
momentum space deformations are possible and collisions can survive
even at very low temperatures. So far, it has not been possible to
completely control this problem from an experimental point of view
and, for this reason, no firm conclusions about superfluidity can
actually be drawn.

Another limitation is related to the hydrodynamic approach:
hydrodynamics can be safely applied only within the limits of validity
of semiclassical approaches, $\Delta \gg \hbar \Omega$, where $\Delta$
is the pairing gap and $\Omega$ is the trapping frequency. Effects
from the finite size and inhomogeneity, governed by the finite trap
frequency $\Omega$, are neglected.  Moreover, the hydrodynamic
formalism has been developed so far only for the case of zero
temperature ($T=0$).

In this article we deal with the excitation spectra in the normal and
superfluid phases of a dilute Fermi gas and we analyze how these
spectra are affected by superfluidity, both in hydrodynamic and
microscopic descriptions.  In order to study excitations similar to
those observed experimentally (the breathing modes) we focused our
attention on the monopole and quadrupole modes. However, while the
breathing modes have been observed for a cigar-shaped gas (and the
radial and axial frequencies have been measured), we restrict our
analysis to a spherical gas for the sake of numerical
tractability. Moreover, while the experiment of \Ref{modes} has been
done for a strongly interacting gas, we treat a weakly interacting
system.
 
We analyze the excitation spectra within a finite-temperature
mean-field approach which provides a microscopic treatment for the
system. The Bogoliubov-de Gennes (BdG) equations \cite{deGennes} are
solved for the ground state and the excitations are treated within the
quasiparticle random-phase approximation (QRPA) \cite{Anderson}. This
approach has already been developed for atomic Fermi gases in
\Ref{BruunMottelson}, where the spin-dipole and the quadrupole modes
have been analyzed. On the other hand, the monopole modes have already
been studied and compared to a schematic model in \Ref{brmon}.

In the present work we want to study systematically the effects
related to the temperature and to the trap frequency of the system. In
particular, we compare our results with the corresponding hydrodynamic
ones in order to check the validity of the semiclassical approach. In
addition to the strength distributions related to the excitation
spectra, we also present the transition densities which can give
important information on nature of the collective modes.

The article is organized as follows. In \Sec{formalism} we briefly
sketch the quantum mechanical and semiclassical formalisms to describe
collective modes in the superfluid phase and in the normal phase in
the collisionless limit.  In \Sec{results} results for the monopole
and quadrupole excitations are shown: the dependence on the
temperature and on the frequency of the trap are studied. In
\Sec{conclusions} we draw our conclusions.

\section{Quantum mechanical and semiclassical formalism}
\label{formalism}
In this section we will briefly review the theoretical description of
collective modes in trapped Fermi gases. As already mentioned in the
introduction, one has to distinguish between quantum mechanical
(``microscopic'') and semiclassical approaches. The fully quantum
mechanical calculation consists in solving the QRPA equations, which
are the small-amplitude limit of the time-dependent BdG equations. At
present such calculations are available only for systems containing up
to $\sim 10^4$ atoms in the case of a spherically symmetric
trap. These conditions are quite far from the experimental ones,
corresponding to particle numbers of $\sim 10^5-10^6$ particles in a
cigar-shaped trap. Up to now, the ``realistic'' conditions can only be
treated within semiclassical approaches. The simplest semiclassical
approach is the hydrodynamic theory. This theory is valid in the
superfluid phase at zero temperature, since the pairing correlations
keep the Fermi surface spherical during the collective motion of the
system. However, hydrodynamics fails at non-zero temperature, unless
the local equilibrium can be ensured by collisions. Since we are
interested in the weakly interacting regime, the collision rate
$1/\tau$ is very small compared to the frequency of the trap. In this
``collisionless'' regime, the Fermi surface becomes locally deformed
during the collective oscillation. This cannot be described by
hydrodynamics, but requires a description in the framework of the
Vlasov equation. The latter is valid in the normal phase, i.e., above
the critical temperature $T_c$. In the intermediate temperature range
$0<T<T_c$, a semiclassical theory is still missing.

\subsection{Quantum mechanical formalism (QRPA)}
The QRPA method has already been applied to trapped Fermi gases in the
weakly \cite{BruunMottelson} as well as in the strongly interacting
regime \cite{OhashiGriffin} and here we will only give a short
summary.

We consider a gas of atoms with mass $m$ in a spherical harmonic trap
with frequency $\Omega$, assuming that the atoms equally occupy two
hyperfine states $\sigma=\uparrow,\downarrow$. Because of the low
temperature and density of the gas, the interaction between the atoms
can be chosen as a zero-range interaction and parametrized by the
s-wave atom-atom scattering length $a$. In order to simplify the
notation, we will express all quantities in harmonic oscillator (HO)
units, i.e., frequencies in units of $\Omega$, energies in units of
$\hbar \Omega$, temperatures in units of $\hbar \Omega / k_B$, and
lengths in units of the oscillator length $l_{HO} = \sqrt{\hbar / (m
\Omega)}$. Furthermore, instead of the scattering length we will use
the coupling constant $g = 4 \pi a / l_{HO}$ as parameter of the
interaction strength.

As mentioned above, the QRPA describes small-amplitude oscillations
around the equilibrium state within the BdG formalism. Therefore, the
first step consists in solving the BdG equations \cite{deGennes}
\begin{equation}
\begin{split}
[H_0 + W(r)] u_{nlm}(\rv) + \Delta(r) v_{nlm}(\rv) 
  &= E_{nl} u_{nlm}(\rv)\,,\\
\Delta(r) u_{nlm}(\rv) - [H_0 + W(r)] v_{nlm}(\rv) 
  &= E_{nl} v_{nlm}(\rv)
\end{split}
\label{bdgeq}
\end{equation}
for the static case. In this way we obtain a set of quasiparticle
energies $E_{nl}$ and wave-functions $u_{nlm}$ and $v_{nlm}$. In
\Eq{bdgeq}, $H_0$ denotes the hamiltonian of the non-interacting HO
minus the chemical potential $\mu$,
\begin{equation}
H_0 = \frac{1}{2} (-\nablav^2 + r^2) - \mu\,,
\end{equation}
while the interaction is accounted for in a self-consistent way
through the Hartree potential $W$ and the pairing field $\Delta$. Due
to spherical symmetry, the wave functions can be written as
\begin{gather}
u_{nlm}(\vek{r}) = u_{nl}(r) Y_{lm}(\theta,\phi)\,,\\
v_{nlm}(\vek{r}) = v_{nl}(r) Y_{lm}(\theta,\phi)\,.
\end{gather}
The quantum numbers $l$ and $m$ are the angular momentum and its
projection, while $n$ numbers different states having the same $l$ and
$m$. In practice, the diagonalization of \Eq{bdgeq} is done in a
truncated harmonic oscillator basis, containing the eigenfunctions of
the trapping potential up to a certain HO energy $E_C =
N_C+\frac{3}{2}$, i.e.,
\begin{equation}
2(n-1)+l \leq N_C\,.
\end{equation}

The self-consistency relates $W$ and $\Delta$ to the wave functions
$u$ and $v$. The mean field $W$ is just proportional to the density,
i.e.,
\begin{equation}
W(\rv) = g \sum_{nl}^{N_C} \frac{2l+1}{4\pi}
  \{v_{nl}^2(r) [1-f(E_{nl})]\\+u_{nl}(r) f(E_{nl})\}\,,
\label{weq}
\end{equation}
where
\begin{equation}
f(E) = \frac{1}{e^{E/T}+1}
\end{equation}
denotes the Fermi function. The Hartree field is independent of the
cutoff $N_C$ if the latter is taken sufficiently large. The
calculation of the pairing field $\Delta$, however, is more
complicated. The zero-range interaction leads to a divergence which in
the case of uniform systems can be regularized in a standard way by
renormalizing the scattering length. This regularization method has
been generalized to the case of trapped systems by Bruun et
al. \cite{BruunCastin} and developed further by Bulgac and Yu
\cite{BulgacYu} and two of the authors \cite{GrassoUrban}. As a
result, the pairing field can be written as
\begin{equation}
\Delta(\rv) = -g_{\mathit{eff}}(r) \sum_{nl}^{N_C}
\frac{2l+1}{4\pi} u_{nl}(r)v_{nl}(r) [1-2f(E_{nl})]\,,
\label{deltaeq}
\end{equation}
with an effective coupling constant $g_{\mathit{eff}}$ which allows to
include the contribution from states beyond the cutoff $N_C$ within
the Thomas-Fermi approximation (TFA). The explicit expression for
$g_{\mathit{eff}}$ reads
\begin{equation}
\frac{1}{g_{\mathit{eff}}(r)} = \frac{1}{g} + \frac{1}{2\pi^2} 
\Big(\frac{k_F(r)}{2}
  \ln \frac{k_C(r) + k_F(r)}{k_C(r) - k_F(r)} - k_C(r) \Big)\,,
\label{geffeq}
\end{equation}
where $k_F$ and $k_C$ denote the local Fermi and cutoff momenta,
respectively:
\begin{gather}
k_F(r) = \sqrt{2\mu-r^2-2W(r)}\,,\\
k_C(r) = \sqrt{2N_C+3-r^2}\,.
\end{gather}

Once the static BdG equations are solved, we can calculate the linear
response of the system to a small time-dependent
perturbation. Following \Ref{BruunMottelson}, we have to compute the
QRPA response function $\Pi$, which is a $4\times 4$ matrix built out
of 16 correlation functions:
\begin{equation}
\Pi(\omega,\rv,\rvp) = \left(\begin{matrix} 
\llangle \hat{\rho}_\uparrow\hat{\rho}_\uparrow\rrangle &
\llangle \hat{\rho}_\uparrow\hat{\rho}_\downarrow\rrangle &
\llangle \hat{\rho}_\uparrow\hat{\chi}\rrangle &
\llangle \hat{\rho}_\uparrow\hat{\chi}^\dagger\rrangle \\
\llangle \hat{\rho}_\downarrow\hat{\rho}_\uparrow\rrangle &
\llangle \hat{\rho}_\downarrow\hat{\rho}_\downarrow\rrangle &
\llangle \hat{\rho}_\downarrow\hat{\chi}\rrangle &
\llangle \hat{\rho}_\downarrow\hat{\chi}^\dagger\rrangle \\
\llangle \hat{\chi}\hat{\rho}_\uparrow\rrangle &
\llangle \hat{\chi}\hat{\rho}_\downarrow\rrangle &
\llangle \hat{\chi}\hat{\chi}\rrangle &
\llangle \hat{\chi}\hat{\chi}^\dagger\rrangle \\
\llangle \hat{\chi}^\dagger\hat{\rho}_\uparrow\rrangle &
\llangle \hat{\chi}^\dagger\hat{\rho}_\downarrow\rrangle &
\llangle \hat{\chi}^\dagger\hat{\chi}\rrangle &
\llangle \hat{\chi}^\dagger\hat{\chi}^\dagger\rrangle
\end{matrix}\right)\,,
\label{pieq}
\end{equation}
with the short-hand notation
\begin{equation}
\llangle\hat{A}\hat{B}\rrangle = -i\int_0^\infty \frac{dt}{2\pi}
  e^{i\omega t} \langle [\hat{A}(t,\rv),\hat{B}(0,\rvp)]\rangle\,,
\end{equation}
where $\langle\rangle$ means the thermal average. The operators
of the normal and anomalous densities, $\hat{\rho}$ and $\hat{\chi}$,
are defined in terms of the field operators $\hat{\psi}$ and
$\hat{\psi}^\dagger$ as follows:
\begin{gather}
\hat{\rho}_\sigma(t,\rv) = \hat{\psi}^\dagger_\sigma(t,\rv)
  \hat{\psi}_\sigma(t,\rv)\,,\label{rhoeq}\\
\hat{\chi}(t,\rv) = \hat{\psi}_\downarrow(t,\rv)
  \hat{\psi}_\uparrow(t,\rv)\,.\label{kappaeq}
\end{gather}

In order to obtain $\Pi$, we first compute the free or unperturbed
response function $\Pi_0$, which is defined analogously to
\Eq{pieq}, but which does not include the effect of interactions
between the quasiparticles. Thus $\Pi_0$ can be obtained by replacing
the field operators $\hat{\psi}$ in \Eqs{rhoeq} and (\ref{kappaeq}) by
\begin{multline}
\hat{\psi}_\sigma(t,\rv) = \sum_{nlm} [b_{nlm\sigma} u_{nlm}(\rv)
  e^{iE_{nl}t}\\ -\sigma b^\dagger_{nlm-\sigma} v^*_{nlm}(\rv)
  e^{-iE_{nl}t}]\,,
\end{multline}
where $\hat{b}$ and $\hat{b}^\dagger$ are annihilation and creation
operators of non-interacting quasiparticles. Inserting the resulting
expressions into \Eq{pieq} and using the relations $\{b_\alpha,
b_\beta\} = \{b^\dagger_\alpha, b^\dagger_\beta\} = 0$, $\{b_\alpha,
b^\dagger_\beta\} = \delta_{\alpha\beta}(1-f(E_\alpha))$, and $\langle
b^\dagger_\alpha b_\beta \rangle = f(E_\alpha) \delta_{\alpha\beta}$,
we obtain explicit expressions for the 16 functions contained in
$\Pi_0$ in terms of the $u$ and $v$ functions and the quasiparticle
energies obtained from \Eq{bdgeq}.

Due to the spherical symmetry of the trap and the rotational
invariance of the interaction, excitations with different angular
momenta do not mix. Therefore it is useful to decompose $\Pi_0$ into
contributions of different angular momenta:
\begin{equation}
\Pi_0(\omega,\rv,\rvp) = \sum_{LM} \Pi_{0L}(\omega,r,\rp)
Y_{LM}(\theta,\phi) Y_{LM}^*(\theta^\prime,\phi^\prime)\,.
\end{equation}

The QRPA response $\Pi_L$ for angular momentum $L$ can now be obtained
from the quasiparticle response $\Pi_{0L}$ by solving the
Bethe-Salpeter integral equation
\begin{multline}
\Pi_L(\omega,r,\rp) = \Pi_{0L}(\omega,r,\rp)\\
 +\int_0^\infty d\rpp r^{\prime\prime 2}\Pi_{0L}(\omega,r,\rpp) G
  \Pi_L(\omega,\rpp,\rp)\,,
\label{qrpaeq}
\end{multline}
where $G$ accounts for the residual interaction between the
quasiparticles:
\begin{equation}
G = \left(\begin{matrix}
  0&g&0&0\\g&0&0&0\\0&0&0&g\\0&0&g&0\end{matrix}\right)\,.
\end{equation}

When calculating the 16 functions contained in $\Pi_{0L}$, one
observes that two of them, namely those related to $\llangle
\hat{\chi}^\dagger \hat{\chi} \rrangle$ and $\llangle \hat{\chi}
\hat{\chi}^\dagger \rrangle$, are divergent for $N_C\to\infty$. This
divergence has the same origin as that of the pairing field. Bruun and
Mottelson \cite{BruunMottelson} therefore suggested to use the same
pseudopotential method as for the regularization of the pairing field
in order to remove the divergence. However, it is not clear how in
their prescription, Eq.\@ (7) in \Ref{BruunMottelson}, the
contribution of states beyond the cutoff $N_C$ can be approximated (as
we did in the case of the pairing field by using the TFA), which is
crucial for having convergence at reasonable values of the cutoff
$N_C$. We therefore propose a simplified prescription: when
calculating $\Pi_{0L}$, we have to restrict the sum to states below
the cutoff, $2(n-1)+l\leq N_C$. To compensate the resulting cutoff
dependence, the interaction in the pairing channel must be replaced by
the effective coupling constant given in \Eq{geffeq}. Thus, we replace
$G$ in \Eq{qrpaeq} by $G_{\mathit{eff}}(\rpp)$, which is defined by
\begin{equation}
G_{\mathit{eff}}(r) = \left(\begin{matrix}0&g&0&0\\g&0&0&0\\
  0&0&0&g_{\mathit{eff}}(r)
\\0&0&g_{\mathit{eff}}(r)&0\end{matrix}\right)\,.
\end{equation}
One can show that, in the case of a uniform system, this simplified
prescription coincides with the pseudopotential method in the limit of
excitations with long wavelengths and low frequencies. We have checked
the convergence of the results using this regularization prescription.

Finally, we have to say how physical quantities of interest can be
extracted from the correlation function $\Pi$. To that end it is
useful to look at the spectral representation
\begin{equation}
\sum_{\sigma\sigmap}
  \llangle\hat{\rho}_\sigma\hat{\rho}_{\sigmap}\rrangle
  = \int d\omega^\prime
  \frac{S(\omega^\prime,\rv,\rvp)}
  {\omega-\omega^\prime+i\varepsilon}\,,
\end{equation}
with
\begin{multline}
S(\omega,\rv,\rvp) = -\frac{1}{\pi}\sum_{\sigma\sigmap}\Im
  \llangle\hat{\rho}_\sigma\hat{\rho}_{\sigmap}\rrangle\\
  = (1-e^{-\omega/T})\sum_{ij}\frac{e^{-E_i/T}}{Z}
  \delta(\omega-E_j+E_i)\\
\times \sum_{\sigma\sigmap}
  \langle i|\hat{\rho}_\sigma(\rv)|j\rangle
  \langle j|\hat{\rho}_{\sigmap}(\rvp)|i\rangle\,,
\label{seq}
\end{multline}
where $|i\rangle$ and $|j\rangle$ are eigenstates of the many-body
hamiltonian with total energies $E_i$ and $E_j$, respectively, and $Z
= \sum_i \exp(E_i/T)$. In the present QRPA formalism Eq. (\ref{seq})
is evaluated using the four upper left elements of the $\Pi$ response
function (\ref{pieq}), obtained with Eq. (\ref{qrpaeq}).

In this paper we will consider excitation operators of the form
\begin{equation}
V_1(t,\rv) \propto r^2 Y_{LM}(\theta,\phi) e^{-i\omega t}\,.
\label{v1eq}
\end{equation}
with $L=0$ (monopole excitations) and $L=2$ (quadrupole
excitations). The corresponding strength function $S_L(\omega)$, which
gives the excitation spectrum, is defined by
\begin{equation}
S_L(\omega) = \int_0^\infty dr r^4 \int_0^\infty d\rp 
  r^{\prime 4} \sum_{\sigma\sigmap}
  S_L(\omega,r,\rp)\,.
\end{equation}
Another interesting quantity is the transition density $\delta\rho =
\rho-\rho_0$, where $\rho_0$ denotes the density in equilibrium and
$\rho$ is the density of the excited system. In the case of zero
temperature, where the stationary system is in the ground state
$|0\rangle$, the transition density for $\omega = E_j-E_0$ is
proportional to
\begin{equation}
\delta\rho(\omega=E_j-E_0,\rv) \propto \sum_\sigma \langle
j|\hat{\rho}_\sigma(\rv)|0\rangle\,.
\end{equation}
In this case, the sum over $i$ in \Eq{seq} reduces to one term
($i=0$), and therefore the transition density can be obtained from
\begin{equation}
[\delta\rho(\omega=E_j-E_0,\rv)]^2 \propto
  \int_{\omega-\delta}^{\omega+\delta} d\omega^\prime
  S(\omega^\prime,\rv,\rv)\,,
\label{td}
\end{equation}
where $\delta$ is supposed to be sufficiently small to avoid that
other states than the selected one ($|j\rangle$) contribute.

\subsection{Superfluid hydrodynamics}
\label{hydrodynamics}
At zero temperature, superfluid hydrodynamics provides the equations
of motion for the density (per spin state) $\rho(t,\rv)$ and the
irrotational collective velocity field $\vv(t,\rv)$ of the superfluid
current (continuity and Euler equations) \cite{CozziniStringari}:
\begin{gather}
\dot{\rho}+\nablav\cdot(\rho\vv)=0\,,
\label{contieq}\\
\dot{\vv} = -\nablav\Big(\frac{\vv^2}{2}
  +\frac{\Vext}{m}+\frac{\muloc}{m}\Big)\,.
\label{eulereq}
\end{gather}
These equations can equally be used for fermionic and bosonic systems,
only the equation of state, relating the local chemical potential
$\muloc$ to the density $\rho$, must be adapted correspondingly. In
the case of weakly interacting fermions, where the density can be
regarded as independent of the pairing gap, this equation of state is
given by the Thomas-Fermi relation
\begin{equation}
\muloc(\rho) = \frac{p_F^2}{2m}+g \rho
  = \frac{\hbar^2 (6\pi^2\rho)^{2/3}}{2m} + g \rho\,.
\label{eos}
\end{equation}
In the static (equilibrium) case, \Eq{eulereq} together with this
equation of state gives immediately the usual Thomas-Fermi equation
for the density profile $\rho_0(\rv)$,
\begin{equation}
\muloc[\rho_0(\rv)]+V_0(\rv) = \mu\,,
\label{hydrostat}
\end{equation}
which is valid in both the normal and the superfluid phase. While the
TFA in the normal phase is valid if $\muloc$ is
much larger than the discrete level spacing of the trapped system
($\hbar\Omega$ in our case), superfluid hydrodynamics requires in
addition that also the pairing gap $\Delta$ is large compared with the
level spacing, which is much more difficult to satisfy.

Since the superfluid velocity field $\vv$ is irrotational, it can be
written as a gradient. In order to establish a connection with
microscopic quantities, we write it in the form
\begin{equation}
\vv(\rv) = \frac{\hbar}{m}\nablav \varphi(\rv)\,.
\end{equation}
where $\varphi$ is related to the phase of the pairing field by
$\Delta(\rv) = |\Delta(\rv)|\exp[2i\varphi(\rv)]$.

In this article we are interested in small-amplitude motion. We
therefore split the density and the external potential into their
equilibrium values and small deviations, $\rho = \rho_0+\delta \rho$
and $V_\mathit{ext} = V_0+V_1$, and expand \Eqs{contieq} and
(\ref{eulereq}) up to linear order in the deviations. In addition, as
we did in the preceding subsection, we will specialize to the case of
a spherically symmetric harmonic trap and use the corresponding HO
units ($\hbar=m=\Omega=1$), i.e., $V_0 = r^2/2$. We know that for an
excitation of the type (\ref{v1eq}) the solution must be of the form
\begin{equation}
\varphi(t,\rv) = \varphi(r)Y_{LM}(\theta,\phi)\exp(-i\omega t)
\end{equation}
and analogous for $\delta\rho$. Furthermore, we are interested in the
eigenmodes of the system, which persist even if $V_1 = 0$. Then
\Eqs{contieq} and (\ref{eulereq}) can be transformed into an
eigenvalue equation for the eigenfrequencies $\omega$ and the
corresponding eigenfunctions $\varphi(r)$,
\begin{equation}
\frac{d\muloc}{d\rho}\Big|_{\rho_0}
  \Big(\frac{1}{r^2}(r^2\rho_0\varphi^\prime)^\prime
  -L(L+1)\varphi\Big)
  =-\omega^2 \varphi\,,
\label{hydrodgl}
\end{equation}
where $f^\prime$ means $df/dr$, and an equation for the transition
density,
\begin{equation}
\delta\rho = -i\omega
  \Big(\frac{d\muloc}{d\rho}\Big|_{\rho_0}\Big)^{-1} \varphi
  = \frac{-i\omega}{r}\rho_0^\prime \varphi\,.
\label{transdens}
\end{equation}

The numerical solution of \Eq{hydrodgl} is not difficult. However, in
the present article we are only interested in the lowest monopole
($L=0$) and quadrupole ($L=2$) modes. For these two modes, the
velocity field $\vv$ is practically linear in $\rv$, and we can thus
obtain a very accurate analytic approximation to the numerical
solution. Let us start with the quadrupole mode $(L=2)$. We insert the
ansatz $\varphi \approx a r^2$ into \Eq{hydrodgl}, multiply the
equation by $\rho_0(r)$ and integrate over $d^3r$. By this integration
the small deviations of the quadratic ansatz from the exact solution
of \Eq{hydrodgl} are averaged out and one thus obtains a very precise
prediction for the frequency. After a lengthy calculation we reproduce
the well-known result
\begin{equation}
\omega_{L=2} = \sqrt{2}\,,
\end{equation}  
which is independent of the interaction.

In a similar way we can find an approximation for the eigenfrequency
of the lowest monopole mode ($L=0$). In this case the function
$\varphi$ has the form $\varphi(r) \approx a - b r^2$. Inserting this
ansatz into \Eq{hydrodgl}, taking the derivative with respect to $r$
in order to get rid of the constant $a$, multiplying by $r$ and
proceeding in the same way as in the case of the quadrupole mode, we
finally obtain
\begin{equation}
\omega_{L=0} = 2\sqrt{1+\frac{3\Eint}{8\Epot}}\,,
\end{equation}
where $\Eint$ and $\Epot$ are the interaction and potential energies,
\begin{equation}
\Eint = \int d^3r g\rho_0^2(\rv)\,,\quad
\Epot = \int d^3r r^2 \rho_0(\rv)\,.
\end{equation}
Contrary to the quadrupole frequency, the monopole frequency depends
on the interaction. Since $\Eint$ is negative, the frequency
$\omega_{L=0}$ is slightly lower than twice the trap frequency,
$2\Omega$. Finally, the ratio of the constants $a$ and $b$, which is
needed in order to compute the transition density $\delta\rho$, can be
determined from the condition that the integral over $\delta\rho$ must
vanish, since the total number of particles stays constant.

\subsection{Vlasov description}
Let us now consider a normal Fermi gas just above $T_c$. In the weakly
interacting limit, $T_c$ is very small as compared with the Fermi
energy, i.e., except for the fact that the system is not superfluid,
we can neglect temperature effects. We will also assume that the
effect of collisions can be neglected. Under this condition the system
cannot come to local equilibrium during the collective motion. In
order to describe this effect, we will use the Wigner function
$f(t,\rv,\pv)$. In equilibrium and within the TFA, this function
simply describes a Fermi sphere:
\begin{equation}
f_0(\rv,\pv) = \Theta(p_F(\rv)-p)\,.
\end{equation}
Out of equilibrium, if the particles do not undergo enough collisions
to restore the isotropic momentum distribution, the local Fermi
surface will assume a more complicated shape. The equation of motion
for the Wigner function is the Vlasov equation
\begin{equation}
\dot{f} = (\nablav V)\cdot(\nablav_p f)
  -\frac{\pv}{m}\cdot(\nablav_r f)\,,
\label{vlasoveq}
\end{equation}
where $V(t,\rv) = \Vext(t,\rv) + g\rho(t,\rv)$ is the total
(external+mean-field) potential and $\nablav_r$ and $\nablav_p$ are
acting in coordinate and momentum space, respectively.

Contrary to the hydrodynamic equations in the superfluid phase, it is
very difficult to solve the Vlasov equation directly. We are
therefore again looking for approximate solutions for the special case
of small-amplitude monopole and quadrupole oscillations in a spherical
harmonic trap. We will employ the ``generalized scaling ansatz''
\cite{RingSchuck}, which has been used with great success to describe
giant resonances in atomic nuclei and which has also been applied to
trapped atomic Fermi gases \cite{Menotti}. In this approach, the
possible deformations of the local Fermi surface are restricted to
quadrupolar shape. Introducing a small displacement field
$\xiv(t,\rv)$, one can write
\begin{equation}
f(t,\rv,\pv) = f_0(\rvp,\pvp)\,,
\label{scaling}
\end{equation}
with
\begin{gather}
\rvp = \rv-\xiv(t,\rv)\,,
\label{scalingr}\\
\pvp = \pv-m\dot{\xiv}(t,\rv)+\nablav_r[\pv\cdot\xiv(t,\rv)]\,.
\label{scalingp}
\end{gather}
The velocity field is then simply given by $\vv = \dot{\xiv}$, and the
last term in \Eq{scalingp} describes the deformation of the Fermi
sphere. For the form of the velocity field we make the same ansatz as
before, i.e.,
\begin{equation}
\xiv(t,\rv) = a \nablav r^2 Y_{LM}(\theta,\phi) e^{-i\omega t}\,,
\end{equation}
with $L = 0$ (monopole mode) or $L = 2$ (quadrupole mode). In analogy
to the procedure in the preceding subsection, we linearize the Vlasov
equation (\ref{vlasoveq}) with respect to $\xiv$, multiply by
$\pv\cdot\xiv^*$ and integrate over $d^3p$ and $d^3 r$. Using
\Eqs{hydrostat} we reproduce after a tedious calculation the results
originally derived in \Ref{Menotti},
\begin{equation}
\omega_{L=0} = 2\Omega \sqrt{1+\frac{3\Ekin}{8\Epot}}\,,\quad
\omega_{L=2} = 2\Omega \sqrt{1-\frac{3\Ekin}{4\Epot}}\,.
\end{equation}
Note that the monopole mode has the same frequency in the normal phase
as in the superfluid phase. This can be understood as follows. If the
displacement field is purely radial ($\xiv\propto \rv$), as it is the
case for the monopole mode, one can see from \Eq{scaling} that the
Fermi surface stays spherical. Therefore hydrodynamics gives the same
frequency as the Vlasov equation. The frequency of the quadrupole mode
in the normal phase, however, is higher than in the superfluid phase
by a factor of approximately $\sqrt{2}$. From \Eq{scaling} one can see
that in this case the Fermi surface gets a quadrupole deformation
perpendicular to the deformation of the density profile in coordinate
space. This deformation costs energy and therefore increases the
frequency of the mode as compared to hydrodynamics.

\section{Results}
\label{results}
In this section we will compare QRPA and semiclassical results for
monopole and quadrupole oscillations in a spherical trap. We are
mainly interested in the limits of validity of superfluid
hydrodynamics, since this theory is widely used in order to analyze
experimental results. For instance, recent experiments showed that the
axial breathing mode in a cigar-shaped trap follows the hydrodynamic
behavior throughout the BCS-BEC crossover, while the radial breathing
mode deviates considerably from the hydrodynamic predictions
\cite{Bartenstein}, especially on the BCS side of the crossover
region. In this experiment the system was still very strongly
interacting even on the BCS side of the crossover (the strongest
deviations happened when $k_F |a|$ was of the order of $2$), such that
our weak-coupling theory (valid for $k_F |a|\ll 1$) cannot directly be
compared to that experiment. Nevertheless, it is clear that the limits
of validity of hydrodynamics should be clarified. It is known that
hydrodynamics works at zero temperature and if the level spacing
$\hbar\Omega$ is much smaller than the gap $\Delta$, but both
conditions are generally not fulfilled in the experiments. Since
experiments cannot be done at zero temperature, it is interesting to
see what kind of temperature effects can arise below the critical
temperature $T_c$. The second condition is also very strong,
especially if the trap is strongly deformed and the transverse trap
frequency is large, and it is therefore important to know up to which
ratio $\hbar\Omega/\Delta$ hydrodynamics can be trusted.

\subsection{Temperature dependence}
In this subsection we will study how the properties of collective
modes change in the small temperature range from zero to the critical
temperature $T_c$.  For this investigation we are using the parameter
set $\mu = 32$ $\hbar\Omega$ and $g = -0.965$ (in HO units). With
these parameters, the number of particles is approximately $17000$ and
the gap in the center of the trap at zero temperature is approximately
$6\,\hbar\Omega$; one can therefore expect that at least at zero
temperature hydrodynamics should work very well.

\begin{figure}
\includegraphics[width=9cm,height=4.5cm]{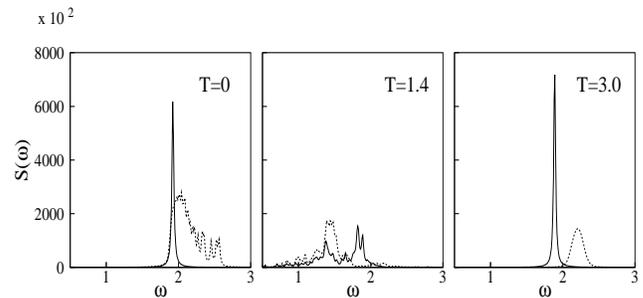}
\caption{Free quasiparticle response (dashed line) and QRPA response
(solid line) of the monopole excitation as a function of the frequency
$\omega$ (in units of the trap frequency $\Omega$), for three
different temperatures: $k_B T=0$, $1.4\,\hbar\Omega$, and $3$
$\hbar\Omega$ (from left to right).
\label{fig:mono}}
\end{figure}

\begin{figure}
\includegraphics[width=9cm,height=4.5cm]{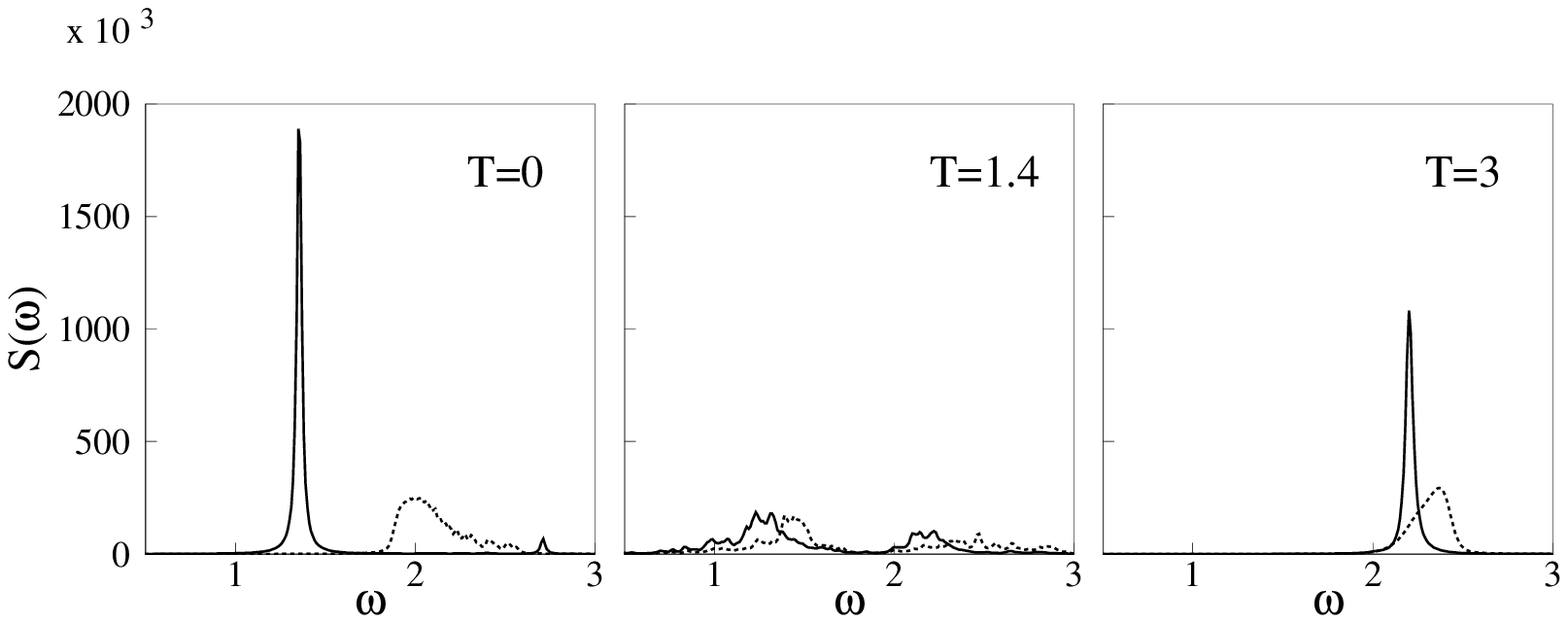}
\caption{Free quasiparticle response (dashed line) and QRPA response
(solid line) of the quadrupole excitation as a function of the
frequency $\omega$ (in units of the trap frequency $\Omega$) for three
different temperatures: $k_B T=0$, $1.4\,\hbar\Omega$, and
$3\,\hbar\Omega$ (from left to right).
\label{fig:quad}}
\end{figure}

\begin{table}
\begin{tabular}{c|c|c|c|c}
      & \multicolumn{2}{c|}{$T = 0$} & \multicolumn{2}{c}{$T > T_c$} \\
      & QRPA      & hydro.           & (Q)RPA     & Vlasov           \\
\hline
$L=0$ & $1.9$     & $1.88$           & $1.9$      & $1.88$           \\
$L=2$ & $1.4$     & $\sqrt{2}$       & $2.2$      & $2.22$           \\
\end{tabular}
\caption{\label{tabt} Frequencies (in units of the trap frequency
  $\Omega$) of monopole ($L=0$) and quadrupole ($L=2$) modes for $\mu
  = 32$ $\hbar\Omega$ and $g = -0.965$ (in HO units) at zero
  temperature and above $T_c$. The QRPA results for $T > T_c$ were
  obtained with $T = 3$ $\hbar\Omega/k_B$.}
\end{table}

In \Figs{fig:mono} and \ref{fig:quad} we show the monopole and
quadrupole response functions, respectively, for three different
values of the temperature. The figures on the left show the response
at zero temperature. The solid lines correspond to the QRPA results
while the dashed lines represent the free quasiparticle response. In
principle, the response function consists of a very large number of
discrete levels.  For the purpose of graphical presentation, these
delta functions must be smeared out, and we therefore introduce a
small imaginary part of $\epsilon=0.015\,\Omega$ in the denominators
of the correlation functions [see \Eq{seq}]. For T=0, the QRPA
quadrupole response shows one single collective peak whose frequency
is very close to that predicted by hydrodynamics (see Table I).  The
QRPA response is completely different from the free quasiparticle
response, which has a broad and almost continuous distribution of
strength between $\sim 1.8\,\Omega$ and $\sim 2.7\,\Omega$.  As has
been realized before \cite{BruunMottelson,OhashiGriffin}, the
threshold of the two-quasiparticle strength is related to the energy
of the lowest-lying quasiparticles which are located near the surface
of the atomic cloud.

In the case of the monopole mode the good agreement between QRPA and
hydrodynamics (Table I) is even more surprising than in the case of
the quadrupole mode, since the frequency of the monopole mode is so
high that it lies in the two-quasiparticle continuum (see dashed line
in \Fig{fig:mono}) and one would therefore expect a certain amount of
Landau damping.

\begin{figure}
\includegraphics[width=9cm,height=4.5cm]{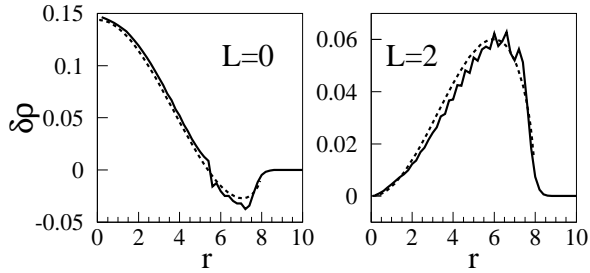}
\caption{Transition densities for the collective monopole (left panel)
  and quadrupole (right panel) modes as a function of $r$ (in units of
  the oscillator length $l_{HO}$), at $T = 0$. Solid and dashed lines
  represent the QRPA and the semiclassical results, respectively.
\label{fig:td}}
\end{figure}

Apart from the study of the frequencies of the collective modes, the
comparison between hydrodynamics and QRPA can be extended also to the
analysis of the character of such modes. We display in \Fig{fig:td}
the transition densities of the two collective modes, which, since the
density profile is known, can be related to the velocity field [see
\Eq{transdens}].  The normalization of the QRPA transition density is
obtained from the integral of the corresponding peak in the strength
function, while that of the semiclassical transition density has been
adjusted to the QRPA one. We see that the simple formulas from
\Sec{hydrodynamics} are in good agreement with the QRPA transition
densities.  However, the QRPA transition densities exhibit small
Friedel-like oscillations, especially near the surface where the gap
is small and the local Fermi surface is therefore relatively sharp.

Let us now consider an intermediate temperature between $0$ and
$T_c$. For the present set of parameters the critical temperature is
$T_c \approx 2.8\,\hbar\Omega/k_B$; we therefore choose $T =
1.4\,\hbar\Omega/k_B \approx T_c/2$. As can be seen in the middle of
\Figs{fig:mono} and \ref{fig:quad}, due to the presence of thermally
excited quasiparticles the free quasiparticle response starts now
already at $\omega = 0$. As a consequence, both the collective
monopole and quadrupole modes become strongly fragmented and
damped. Qualitatively, this strong Landau damping at temperatures
between zero and $T_c$ could be related to the damping mechanism which
is responsible for the experimentally observed damping of breathing
modes on the BCS side of the BEC-BCS crossover
\cite{Bartenstein}. Interesting is also the double-peak structure
which can be seen in the quadrupole response, as if there were two
damped modes, one corresponding to the hydrodynamic mode and another
one corresponding to the quadrupole mode in the collisionless normal
phase (see below). This can be interpreted in the sense of the
two-fluid model \cite{Leggett}, which states that between $T = 0$ and
$T = T_c$ the system effectively behaves as if it consisted of a
mixture of normal and superfluid components.

Now we increase the temperature further to $T = 3\,\hbar\Omega/k_B$,
which lies slightly above $T_c$, i.e., the system reaches the normal
phase, but still the temperature is very low compared with the Fermi
energy. In the normal phase, the BdG equations become identical to the
usual Hartree-Fock equations, and the QRPA becomes equal to the usual
RPA. In the case of the monopole mode (right panel of \Fig{fig:mono}),
the QRPA response is almost identical to that at zero temperature
(left panel of \Fig{fig:mono}), although the free quasiparticle
response is quite different. Again there is one collective mode having
the same frequency as at $T=0$. This is not very surprising. As
mentioned in the preceding section, the Vlasov equation predicts the
same frequency as superfluid hydrodynamics, since in the case of the
monopole mode there is no deformation of the local Fermi surface. This
is different in the case of the quadrupole mode (right panel of
\Fig{fig:quad}).  Also here a collective mode reappears, but it is
situated at a different frequency than at zero temperature. The higher
frequency in the normal phase compared with the superfluid phase is
due to the Fermi surface deformation and is well described by the
Vlasov equation (cf. \Tab{tabt}).

\subsection{Dependence on the size of the system}
Let us now investigate the importance of the discrete level spacing at
zero temperature. In the case without superfluidity, the semiclassical
$\hbar\to 0$ limit (TFA in equilibrium and the Vlasov equation in the
dynamical case) is known to work very well if the number of particles
is sufficiently large. The reason is very simple: The only
dimensionless parameter on which corrections can depend is
$\hbar\Omega/\mu$, which becomes very small for large numbers of
particles. In the current experiments involving $\sim 10^5-10^6$ atoms
this type of corrections is completely negligible. For our study we
choose, as in the preceding subsection, a chemical potential of
$\mu=32\,\hbar\Omega$. This is large enough to make these corrections
small, and the numerical calculations are still tractable.  The
corresponding numbers of atoms lie between $\sim 14000$ and $\sim
17000$ depending on the chosen values of the coupling constant $g$ due
to the Hartree field (see \Tab{tabg}).

\begin{table}
\begin{tabular}{c|c|c}
$g$      & $N$          & $\Delta(0)$\\
\hline
$-0.965$ & $ 16500$ & $ 6.0$   \\
$-0.8$   & $ 15000$ & $ 2.9$ \\
$-0.7$   & $ 14300$ & $ 1.4$ \\
$-0.636$ & $ 13900$ & $ 0.7$
\end{tabular}
\caption{\label{tabg} Chosen values of the coupling constant $g$
(first column; in HO units) and corresponding results for the number
of particles, $N$ (second column), and for the gap at the center of
the trap, $\Delta(0)$ (third column; in units of $\hbar\Omega$). The
remaining parameters were fixed to $\mu = 32$ $\hbar\Omega$ and $T =
0$.}
\end{table}

In the case of superfluidity, however, another dimensionless parameter
becomes important, which is $\hbar\Omega/\Delta$. Since in the BCS
phase $\Delta\ll\mu$, this parameter is not necessarily small even if
the number of particles is very large. In order to study the validity
of hydrodynamics as a function of $\hbar\Omega/\Delta$, we change
$\Delta$ by varying the coupling constant $g$ between $-0.636$ and
$-0.965$ (in HO units). As a measure for $\Delta$ we take its value at
the center of the trap, $\Delta(0)$. The values of $\Delta(0)$
corresponding to the different coupling constants are listed in
\Tab{tabg}.

We are now going to analyze the finite-size effects on the quadrupole
response function by using the different values of the coupling
constant listed in \Tab{tabg}. Note that, since we are using HO units,
changing the coupling constant $g\propto a/l_{HO}$ is equivalent to
changing the oscillator length $l_{HO}$ and thus the radius of the
cloud $R = \sqrt{2\mu/\hbar\Omega}\, l_{HO}$. Anyway, as argued above,
the important parameter for finite-size effects is the ratio
$\hbar\Omega/\Delta(0)$ and not the cloud size itself.

\begin{figure}
\includegraphics[width=9cm,height=4.5cm]{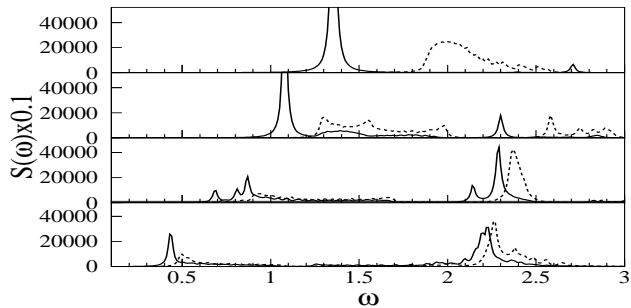}
\caption{Unperturbed response (dashed line) and QRPA response (solid
  line) of the quadrupole excitation as a function of the frequency
  $\omega$ (in units of the trap frequency $\Omega$) for $T=0$ and
  $\mu=32\,\hbar\Omega$ and four different values of the coupling
  constant: $g = -0.965$, $g=-0.8$, $g=-0.7$, and $g=-0.636$ (in HO
  units; from top to bottom).
\label{fig:pairint}}
\end{figure}

For the strongest coupling, $g = -0.965$ (in HO units), the central
value of the gap, $\Delta(0)$, is large compared with $\hbar\Omega$,
and hydrodynamics works almost perfectly at zero temperature, as we
have already seen in the preceding subsection. \Fig{fig:pairint}
shows, from top to bottom, the evolution of the quadrupole response at
$T=0$ for decreasing coupling constant $g$, i.e., for increasing
importance of the discrete level spacing. Besides the QRPA response
(solid lines), we also show the free quasiparticle response (dashed
lines).  For $g = -0.8$ (in HO units), the gap at the center is still
larger than $\hbar\Omega$ by a factor of three, but now we find
considerable deviations of the QRPA response from the hydrodynamic
result. Since the free quasiparticle response is now shifted to lower
frequencies, the hydrodynamic mode becomes fragmented, which
experimentally would show up as damping effect, and its frequency
($\omega\approx 1.1\,\Omega$) lies below the hydrodynamic prediction
($\sqrt{2}\,\Omega$). For $g = - 0.7$ and $g = - 0.636$ (in HO units),
the central value of the gap is comparable to $\hbar\Omega$ and it is
clear that hydrodynamics must fail.  Indeed, the QRPA response becomes
more and more similar to the free quasiparticle response which in the
case of weak pairing looks very different from the strong-pairing
case. The double-peak structure is a consequence of the two types of
transitions which are allowed by the selection rules of the harmonic
oscillator, i.e., transitions inside an oscillator shell ($\delta N =
0$, where $N$ denotes the number of oscillator quanta) and transitions
with $\delta N = 2$. As the interaction decreases, the strength of the
$\delta N=0$ transitions becomes less important while the $\delta N =
2$ transitions become stronger. This can be understood from the fact
that in the limit of a noninteracting harmonic oscillator without
pairing ($g\to 0$) the $\delta N=0$ transitions are forbidden by Pauli
principle and only the $\delta N = 2$ transitions survive. In this
limit the response has a single peak at $\omega = 2\Omega$, in exact
agreement with the prediction from the Vlasov equation
\footnote{In the case of a non-interacting harmonic oscillator, as
well as in the case of a harmonic oscillator with separable
quadrupole-quadrupole interaction, the Vlasov equation is known to
reproduce exactly the quantum mechanical solution.}.
In the semiclassical language, one can say that in this case the
pairing is too weak to restore the spherical shape of the Fermi sphere
during the oscillation, and therefore one finds the normal
collisionless frequency instead of the hydrodynamical one.

\section{Summary and conclusions}
\label{conclusions}
In this article we have studied the properties of collective monopole
and quadrupole modes in superfluid Fermi gases in the BCS phase ($k_F
|a|\ll 1$, $a < 0$) in a spherical harmonic trap. Having briefly
recalled the quantum mechanical and semiclassical formalisms (QRPA,
hydrodynamics, Vlasov equation), we presented numerical results and
compared the different formalisms. Our main interest was focused on
two types of effects: temperature and finite-size effects. Both cannot
be treated within the semiclassical approaches available in the
present literature, and they can therefore only be studied in the
framework of the fully microscopic QRPA formalism.

In the case of a sufficiently large system (large meaning
$\Delta\gg\hbar\Omega$), superfluid hydrodynamics can be used to
describe the properties of collective modes at zero temperature. Our
results confirm earlier findings \cite{BruunMottelson} which show that
already for parameters which lead to $\Delta(0) = 6\hbar\Omega$ the
extremely simple theory of superfluid hydrodynamics is in almost
perfect agreement with the numerically heavy QRPA method. This is not
only true for the frequencies, but also for the transition densities,
i.e., the velocity fields associated with the collective
modes. However, experiments can never be done at zero temperature. The
critical temperature $T_c$ being extremely low, it is clear that
already at very low temperatures between $0$ and $T_c$ the properties
of the collective modes must undergo dramatic changes. This is evident
if the hydrodynamic frequency ($T = 0$) is different from that in the
collisionless normal phase ($T = T_c$), like in the case of the
quadrupole mode. In the case of the monopole mode we also find a
strong temperature dependence, although its frequency at $T = 0$ is
the same as at $T = T_c$. In the intermediate temperature range
between $0$ and $T_c$ the collective modes exhibit strong Landau
damping. When the critical temperature is reached, the damping
disappears and the collective modes can be very well described by the
semiclassical Vlasov equation within the generalized scaling
approximation.

It is interesting to compare these temperature effects with those
found previously in the case of the twist mode \cite{Twist}, which is
an excitation where the upper hemisphere rotates against the lower
one. Near $T_c$, the behavior is rather similar: At $T = T_c$ the
twist mode is a collective mode which can be described by the
generalized scaling approximation to the Vlasov equation and whose
frequency is slightly higher than the trap frequency. If the
temperature is lowered, the twist mode becomes strongly damped, like
the quadrupole and monopole modes. However, an important qualitative
difference appears near zero temperature. Since the velocity field of
the twist mode cannot be written as a gradient, the twist mode
disappears completely at zero temperature, whereas the quadrupole and
monopole modes have an irrotational velocity field and they reappear
at zero temperature as hydrodynamic modes.  In the case of the twist
mode, the disappearance of the $1/\omega$ weighted integrated strength
could be well described within a rather simple two-fluid model
\cite{Twist,Twofluid}. It remains to be studied if a generalization of
the two-fluid model to the dynamical case can also explain the damping
of the quadrupole and monopole modes and the two-peak structure in the
quadrupole response function at temperatures between $0$ and $T_c$.

In addition to temperature effects, we studied how the properties of
the quadrupole mode change at zero temperature when the condition for
the validity of the hydrodynamic approach, $\Delta\gg\hbar\Omega$, is
no longer satisfied.  For parameters leading to $\Delta(0) \approx
3\hbar\Omega$ the QRPA already shows considerable deviations from the
hydrodynamic theory. In the case of the quadrupole mode, the frequency
for these parameters is found to be lower by $20\%$ than the
hydrodynamic prediction, and a certain fragmentation of the excitation
spectrum (i.e., damping of the collective mode) can be observed. If
$\Delta(0) \approx \hbar\Omega$, the hydrodynamic mode has more or
less disappeared. At the same time, a fragmented strength appears in
the excitation spectrum near the frequency of the collective
quadrupole mode in the normal collisionless phase.

These results should be kept in mind when frequencies of collective
modes measured in experiments with strongly deformed traps are
compared with the hydrodynamic predictions. Due to the strong
deformation, the radial trap frequency $\Omega_r$ is often much higher
than the axial one, $\Omega_z$. Even in the case of strong pairing,
the gap might be of the order of, say, $3\,\hbar\Omega_z$, and
considerable deviations from hydrodynamics are possible.
%


\begin{thebibliography}{*}
%
\bibitem{bec} M.H. Anderson, et al., Science \textbf{269}, 198 (1995);
  K.B. Davis, et al., Phys. Rev. Lett. \textbf{75}, 3969 (1995);
  C.C. Bradley, et al., Phys. Rev. Lett. \textbf{75}, 1687 (1995).
\bibitem{fer} M.W. Zwierlein, et al., Phys. Rev. Lett. \textbf{91}, 
250401 (2003) 
\bibitem{expa} K.M. O'Hara, et al., Science \textbf{298}, 2179 (2002).
\bibitem{gap}  C. Chin, et al., Science \textbf{305}, 1128 (2004).
\bibitem{modes} J. Kinast, et al., Phys. Rev. Lett. \textbf{92}, 150402
  (2004).
\bibitem{Menotti} C. Menotti, P. Pedri, and S. Stringari, Phys. Rev. Lett. 
\textbf{89}, 250402 (2002).
\bibitem{Stringari} S. Stringari, Europhys. Lett.\textbf{65}, 749 (2004).
\bibitem{baranov} M.A. Baranov and D.S. Petrov, Phys. Rev. A \textbf{62},
  041601 (2000).
\bibitem{deGennes} P.-G. de Gennes, \textit{Superconductivity of Metals and 
  Alloys} (Benjamin, New York, 1966).
\bibitem{Anderson} P.W. Anderson, Phys. Rev.\textbf{112}, 1900 (1958).
\bibitem{BruunMottelson} G.M. Bruun and B.R. Mottelson, Phys. Rev. Lett.
  \textbf{87}, 270403 (2001).   
\bibitem{brmon} G.M. Bruun, Phys. Rev. Lett.
  \textbf{89}, 263002 (2002).   
\bibitem{OhashiGriffin} Y. Ohashi and A. Griffin, preprint
  cond-mat/0503641 (2005). 
\bibitem{BruunCastin} G. Bruun, Y. Castin, R. Dum, and K. Burnett,
  Eur. Phys. J. D \textbf{7}, 433 (1999).
\bibitem{BulgacYu} A. Bulgac and Y. Yu, Phys. Rev. Lett. \textbf{88},
  042504 (2002).
\bibitem{GrassoUrban} M. Grasso and M. Urban, Phys. Rev. A \textbf{68},
  033610 (2003).
\bibitem{CozziniStringari} M. Cozzini and S. Stringari, Phys. Rev. Lett.
  \textbf{91}, 070401 (2003).
\bibitem{RingSchuck} P. Ring and P. Schuck, \textit{The Nuclear Many-Body 
Problem} (Springer-Verlag, Berlin, 1980).
\bibitem{Bartenstein} M. Bartenstein, et al., Phys. Rev. Lett.
\textbf{92}, 203201 (2004). 
\bibitem{Leggett} A.J. Leggett, Phys. Rev. \textbf{140}, A 1869 (1965);
  Phys. Rev. \textbf{147}, 119 (1966). 
\bibitem{Twist} M. Grasso, M. Urban, and X. Vi{\~n}as, Phys. Rev. A
  \textbf{71}, 013603 (2005).
\bibitem{Twofluid} M. Urban, Phys. Rev. A \textbf{71}, 033611 (2005).
%
\end{thebibliography}
\end{document}